\documentclass[11pt,english,11pt,twosides]{article}
\usepackage[T1]{fontenc}
\usepackage[latin9]{inputenc}
\usepackage{amsthm}
\usepackage{amsmath}
\usepackage{graphicx}
\usepackage{amssymb}
\usepackage[authoryear]{natbib}

\makeatletter

\providecommand{\tabularnewline}{\\}

 \usepackage{asp2010}
\numberwithin{equation}{section}
\numberwithin{figure}{section}
\newcommand{\lyxaddress}[1]{
\par {\raggedright #1
\vspace{1.4em}
\noindent\par}
}

\makeatother

\usepackage{babel}

\begin{document}
\global\long\def\Mo{M_{\odot}}
\global\long\def\Mbh{M_{\bullet}}
\global\long\def\Ms{M_{\star}}
\global\long\def\SgrA{\mathrm{Sgr\, A^{\star}}}
 \global\long\def\Ns{N_{\star}}

\title{Key questions about Galactic Center dynamics}

\author{Tal Alexander}

\

\lyxaddress{\affil{Department of Particle Physics and Astrophysics, Faculty
of Physics, Weizmann Institute of Science, POB 26, Rehovot 76100,
Israel}}
\begin{abstract}
I discuss four key questions about Galactic Center dynamics, their
implications for understanding both the environment of the Galactic
MBH and galactic nuclei in general, and the progress made in addressing
them. The questions are (1) Is the stellar system around the MBH relaxed?
(2) Is there a {}``dark cusp'' around the MBH? (3) What is the origin
of the stellar disk(s)?, and (4) What is the origin of the S-stars? 
\end{abstract}

\section{Introduction: the dynamical components of the GC}

\label{s:intro}

The Galactic Center (GC) is a uniquely accessible laboratory for studying
the dynamics of stars and gas in the vicinity of a massive black hole
(MBH). Commonly assumed theoretical paradigms for interpreting MBHs---key
players in many fields of astrophysics---are strongly challenged by
the observations of the GC. In this short review I discuss four key
questions about Galactic Center dynamics, their implications, and
the progress made in addressing them.

\begin{figure}[t]
\includegraphics[clip,width=1\textwidth]{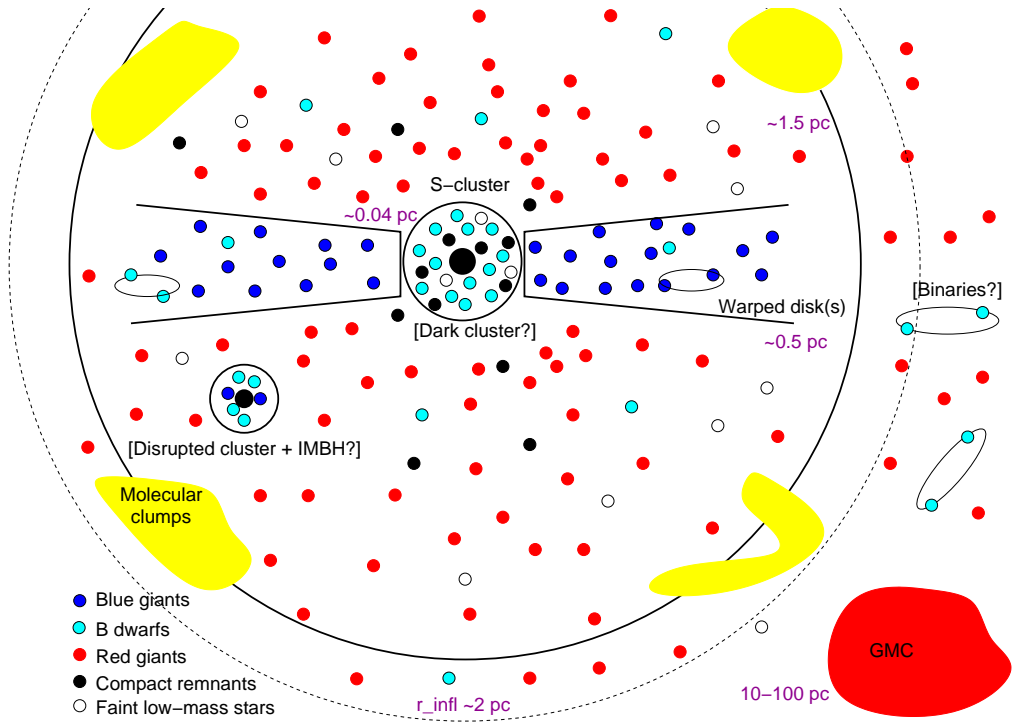}\caption{\label{f:GC}A schematic depiction, not to scale, of the various dynamical
components that are observed in the GC around SgrA$^{\star}$, or
are hypothesized to exist there (these are marked by {[}$\ldots${]},
see section \ref{s:intro}). }

\end{figure}

It useful to set the stage by showing the dynamical components of
the GC in schematic form (Figure \ref{f:GC}). The radius of dynamical
influence of the MBH extends to $\sim2$ pc. Beyond that lies the
central star-forming region of the Galaxy on the $100-200$ pc scale,
which is composed of a mixed population of old low-mass and young
massive stars (among then presumably many binaries)---evidence of
continuous star formation (SF) \citep{fig+04}. It also includes massive
objects such as giant molecular clouds (GMCs) \citep{oka+01} and
stellar clusters \citep{fig+99}. Just inside the radius of influence,
at a distance of $\sim1.5$ pc, lies a ring of less massive molecular
clumps (the circum-nuclear disk, CND), which delineates a central
region with very little gas. The \emph{observed} stellar population
in the central $\sim0.5$ pc is composed of red and blue giants, and
lower-mass main sequence (MS) stars (the faintest currently observed
are B dwarfs). It is assumed that there are many more fainter, yet
unobserved lower-mass main sequence stars there, as well as compact
remnants: white dwarfs (WD), neutron stars (NS) and stellar mass black
holes (BHs). Some over-densities in the stellar distribution in the
inner parsec have been interpreted as the dissolving cores of inspiralling
clusters held together by an intermediate mass BH (IMBH) \citep[e.g.][]{mai+04}.
While the red giants and lower-mass B-dwarfs are isotropically distributed,
the ${\cal O}(100)$ blue giants are concentrated in one (or perhaps
two) coherently-rotating warped disks, which extend inward to a sharp
inner cut-off at $\sim0.04$ pc ($\sim1"$). The inner arcsecond harbors
a spectroscopically and dynamically distinct population of $\sim40$
B-dwarfs on isotropic orbits (the S-cluster).

\section{Is there a relaxed stellar cusp around the MBH?}

\label{s:relax}

The most basic, and arguably the most important question about the
dynamical state of the GC, is whether it is relaxed or not. An unrelaxed
system reflects its particular formation history, which likely varies
substantially from galaxy to galaxy. In contrast, the properties of
a relaxed system can be understood and modeled from first principles,
independently of initial conditions. Lessons learned from a relaxed
GC can then be extrapolated to other relaxed galaxies. For example,
the hypothesis that such an extrapolation is valid is crucial for
understanding the dynamics of extra-galactic gravitational wave (GW)
sources and predicting their rates, since the low-mass MBH in the
GC is the archetype of extra-galactic targets for the planned space-borne
GW observatory LISA \citep{lisa}.

\subsection{Theoretical expectations}

The Galactic Center, like other galactic nuclei with low-mass MBHs,
is expected to be dynamically relaxed. This follows from the observed
correlation between the MBH mass $ $$\Mbh$ and the typical velocity
dispersion of the spheroid of the host galaxy, $\Mbh\!\propto\!\sigma^{\beta}$
with $4\!\lesssim\!\beta\!\lesssim5$ \citep[the $\Mbh/\sigma$ relation, ][]{fer+00,geb+00}.
To see this, assume for simplicity $\beta\!=\!4$ (a higher value
only reinforces this conclusion). The MBH radius of influence $r_{h}\!\sim\! G\Mbh/\sigma^{2}\!\propto\!\!\Mbh^{1/2}$
encompasses a stellar mass of order $\Mbh$, so that the number of
stars enclosed there is $N_{h}\!\sim\! M_{\bullet}/\Ms$, where $\Ms$
is the typical stellar mass, and the mean stellar density is $\bar{n}_{h}\!\sim\! N_{h}/r_{h}^{3}\!\propto\! M_{\bullet}^{-1/2}$.
The simple {}``$nv\Sigma$'' estimate of the rate of gravitational
encounters then implies that the 2-body relaxation rate is $T_{R}^{-1}(r_{h})\!\sim\!\bar{n}_{h}\sigma(G\Ms/\sigma^{2})^{2}\!\propto\!\Mbh^{-5/4}$
(note also for future reference that $T_{R}^{-1}\!\propto\!\Ms^{2}N_{\star}$).
More rigorous estimates yield for the Galactic MBH ($\Mbh\!\simeq\!4\times10^{6}\,\Mo$,
\citealt{eis+05,ghe+05}), $T_{R}\!\sim\!\mathrm{few}\,\mathrm{Gyr}\!<t_{H}$
(the Hubble time) and $\bar{n}_{h}\!\sim\! O(10^{5}\,\mathrm{pc^{-3}})$.
As argued below, the density in a relaxed stellar cusp near a MBH
is orders of magnitude higher still. Since $T_{R}\!\propto\!\Mbh^{-5/4}$,
MBHs with $\Mbh\!\lesssim\!10^{7}\,\Mo$ are expected to lie in relaxed
high density cusps. 

By a coincidence of technology, this also happens to be the MBH mass
range that LISA is sensitive to. This is why GC dynamics are so relevant
for extra-galactic GW sources, in spite of the fact that the chances
of detecting GW emission from the GC itself are small \citep{fre03}.

Relaxed stellar systems around MBHs are expected to settle into a
centrally concentrated, (formally) diverging density distribution---a
cusp. This can easily be seen in the case of a single mass population,
which relaxes to an $r^{-\alpha}$ cusp with $\alpha\!=\!7/4$ \citep{bah+76},
since the gravitational orbital energy gained by the system when stars
are destroyed near the MBH is conserved as it is shared and carried
outward by the remaining stars at a rate $\dot{E}(r)\!\sim\! E(r)N(<r)/T_{R}\propto r^{-1}r^{3-\alpha}/r^{\alpha-3/2}\!=\! r^{7/2-2\alpha}\!=\mathrm{const}$
\citep{bin+87}. When the system includes a spectrum of masses, $M_{L}\le\!\Ms\!\le\! M_{H}$,
the approach toward equipartition by 2-body interactions decreases
the specific kinetic energy of the high-mass stars, while that of
the low-mass stars increases. As a result, the high-mass stars sink
and concentrate in the center on the faster dynamical friction timescale
$T_{\mathrm{df}}\sim T_{R}\left\langle \Ms\right\rangle /M_{H}$,
while the low-mass stars float out \citep{spi87}. The lifespans of
the hot massive stars in the GC are much shorter than the dynamical
friction and relaxation timescales, and therefore these dynamical
processes can significantly affect only longer-lived lower mass, faint
stars, and compact remnants. In particular, stellar mass BHs ($\Ms\sim10\,\Mo$),
which are substantially more massive than any other long-lived species,
are expected to form a dense inner {}``dark cusp'' with a very steep
inner concentration ($\alpha>2$) \citep{ale+09}.

These general theoretical considerations logically lead to two conclusions:
(1) Relaxed systems around MBHs are cusps. (2) Systems without a cusp
(e.g. with a flat density core) are not relaxed.

\subsection{The observed stellar distribution in the GC}

Attempts to characterize the stellar distribution around the Galactic
MBH, first by the integrated light and later by star counts, have
a long history of conflicting results \citep[see e.g. review by][]{gen+94}.
Stellar surface number density maps of the entire stellar population
above the detection threshold (i.e. including both young and old stars)
unambiguously indicate a somewhat shallower cusp than expected, but
one still broadly consistent with the predicted relaxed cusp in the
GC \citep[e.g.][]{sch+07}.

Very recently this picture was overturned with the addition of newly
available stellar classifications for the stars around the MBH, using
narrow band photometry or spectroscopy (Bartko et al. this volume;
Buchholz et al. \citealt{buc+09}; Do et al. \citealt{do+09}; also
in this volume). These observations reveal that the cusp is mostly
or solely composed of massive young stars, whereas the old population
exhibits a core inside $\sim0.5$ pc, or perhaps even a central depletion.

\subsection{Interpretation and implications}

The observed old stars ($K\lesssim17$ mag) are all red giants. A
key assumption in interpreting their surface density distribution
as evidence for the absence of a relaxed cusp is that the giants (typically
$\sim0.01$ of an old stellar population) faithfully track the distribution
of the overall old population. 

The observed core in the giant distribution could still be reconciled
with a relaxed old cusp if some selective mechanism preferentially
destroys giants, or rejuvenates them to appear as hot stars. The latter
would also naturally explain why the inner cusp of young stars appears
to seamlessly continue the old cusp outside the inner $\sim0.5$ pc.
Such destruction and rejuvenation models, which were originally studied
in some detail as possible explanations for the existence of hot stars
in the central parsec \citep[see e.g. review by][]{ale05}, have been
since abandoned in favor of \emph{in situ} SF (section \ref{s:disk}).
It is also unlikely that such processes (tidal heating, \citealt{ale+03a};
envelope stripping by star-giant collisions, \citealt{ale99,bai+99})
can be effective outside the central $\sim0.1$ pc. An extremely massive
cluster of stellar mass BHs born locally from a top-heavy, continuously
forming stellar population (section \ref{s:disk}) could conceivably
destroy the giant progenitors while still on the main sequence throughout
the central $\sim0.5$ pc (Davies et al., this volume). However, the
required mass of this dark cluster exceeds the dynamical limits on
the stellar mass around the Galactic MBH, and is inconsistent with
the {}``drain limit'' (a conservative upper limit on the steady
state number of stellar mass BHs that can survive rapid mutual scattering
into the MBH, \citealt{ale+04}). At this time none of the proposed
selective destruction or rejuvenation mechanisms can plausibly reconcile
the giant core with an old main-sequence relaxed cusp. 

The alternative, that the giants do trace the old population, that
there is no cusp, and that the GC is unrelaxed, can perhaps be the
result of a major perturbation that ejected the stars from the GC
({}``cusp scouring'') sometime in the past. This would increase
the local relaxation time in the center beyond the Hubble time. The
GC would then still be away from equilibrium, slowly returning to
steady state by 2-body stellar relaxation (Merritt, this volume).
Such a destructive event could be a major galactic merger involving
the coalescence of the two MBH \citep{mil+02}. It should be noted
however that a major merger is not \emph{required} to explain the
growth of the low-mass Galactic MBH (it could grow by the direct accretion
of gas and stars, \citealt{fre+02}), and neither are there any other
clear indications of a such a merger in the past apart for the core
in the giant distribution.

Several important implications follow from the absence of a cusp,
if that is indeed the case. In addition to the fact that the properties
of such an unrelaxed core must depend on the details of the core-scouring
event, the much lower density of stars around the MBH imply a much
lower rate of star-star and star-MBH interactions, and in particular
tidal disruption events and GW from extreme mass-ratio inspiral events. 

While a slowly evolving, unrelaxed core could explain the observed
density distribution, this scenario is not without its problems. The
main question is whether the system evolves passively, and on the
slow stellar 2-body relaxation time. There are strong reasons to suspect
that neither these assumptions is correct. 

The latest SF episode in the central $\sim0.5$ pc formed ${\cal O}(100)$
very massive stars that will leave behind stellar BHs \citep{pau+06}.
There are also indications of a previous SF episode ${\cal O}(10^{8}\,\mathrm{yr})$
ago \citep{kra+95}. On average continuous SF in the GC is the best-fit
model to the observed stellar population (\citealt{ale+99a}; Baumgardt,
this volume). Assuming for simplicity that a 100 stellar BHs of mass
$\Ms\!=\!10\,\Mo$ are formed every $10^{8}$ yr (this is consistent
with the ${\cal O}(10^{4})\,\Mo$ total mass of the progenitor gas
disk, \citealt{nay+06}), then $N_{\star}\sim10^{4}$ stellar BHs
are expected to have accumulated in the central $\sim0.5$ pc over
a Hubble time. This implies a short two-body relaxation time of only
$T_{R}\sim Q^{2}P/\left[2\pi\Ns\log Q\right]\sim3\times10^{9}$ yr,
where $Q\equiv M_{\bullet}/M_{\star}$. 

\begin{figure}[t]
\noindent \begin{centering}
\includegraphics[width=0.85\textwidth]{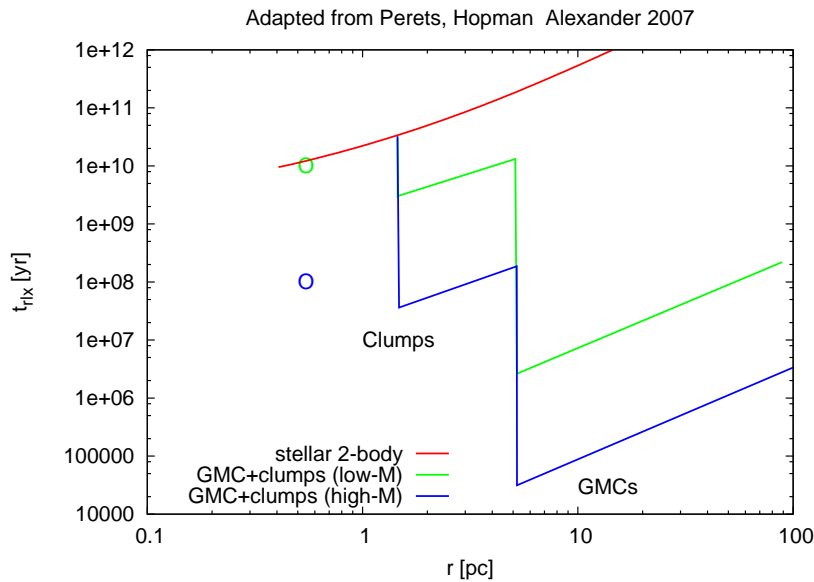}
\par\end{centering}

\caption{\label{f:MPs}Th,e short relaxation time induced by massive perturbers
(mainly GMCs) in the central $100$ pc of the GC \citep{per+07}.
The two bottom sets of curves represent different assumptions about
the MP masses. Inside the inner $\sim1.5$ pc the local perturbers
are stars, but the effects of the gas clumps in the CND extend well
within the inner parsec (circles) and can decrease the relaxation
time well below the Hubble time. }

\end{figure}

In addition, relaxation in the GC on the $\gtrsim1$ pc scale is most
probably dominated by massive perturbers, primarily giant molecular
clouds (GMCs), rather than by stars \citep{per+07}. This is because
$T_{R}^{-1}\propto NM^{2}$ (section \ref{s:relax}), and so the ${\cal O}(100)$
GMCs observed on the l0--$100$ pc scale with up to $\sim10^{7}\,\Mo$,
can decrease the relaxation time by many orders of magnitude. Closer
to the MBH, on the 1--2 pc scale, the less massive CND gas clumps
can decrease the relaxation time well below the Hubble time. The effect
of the CND is reduced, but still very substantial, even on the $\sim0.5$
pc scale (Figure \ref{f:MPs}). The existence of the S-cluster may
be evidence of rapid relaxation by massive perturbers (section \ref{s:Sstars}).

In order for an unrelaxed core to persist in the face of rapid internal
(stellar BHs) and external (CND) perturbations, it is necessary to
assume that the core was formed recently in cosmological terms, or
that both the recent SF activity and the presence of the CND are atypical.
The latter option seems unlikely in view of evidence of continuous
SF and GMC creation and breakup in the GC on all scales \citep{fig+04,per+08c}.
Either of these two explanations implies that we are observing the
GC today at a special epoch in its evolution.

\section{Is there a dark cusp around the MBH?}

\label{s:dark}

A dense, strongly mass-segregated cusp of stellar BHs is expected
near the MBH if the GC is relaxed, and even in non-equilibrium core
models, the reformation of the BH cusp is rapid, although it does
not reach densities as high as predicted for a relaxed system (Merritt,
this volume). Such dark nuclear clusters are expected to play a crucial
role in the generation of extra-galactic GW signals. However, their
existence is yet unconfirmed. Direct detection of the dark cusp in
the GC, for example by gravitational lensing \citep{ale+01c,cha+01a}
or by X-ray emission from accretion \citep{pes+03}, is very difficult.
Dynamical upper limits on the dark distributed mass within the S-cluster
are still at least two orders of magnitude higher than expected from
theoretical constraints \citep{gil+09}. At this time, the most promising
approach to detect them appears to be through their dynamical interactions
with other stars, and in particular by the mechanism of resonant relaxation
(RR) \citep{rau+96,hop+06a}. 

RR occurs when the gravitational potential has approximate symmetries
that restrict orbital evolution (e.g. fixed ellipses in a Keplerian
potential; fixed orbital planes in a spherical potential). In such
cases the perturbations on a test star are no longer random, but correlated,
leading to coherent ($\propto\! t$) torquing of the orbital angular
momentum $\mathbf{J}$ on times shorter than the coherence time $t_{\omega}$,
while the symmetries hold. On longer timescales, coherence is lost
as the orbits slowly change by processes such as in-plane precession
due to the enclosed mass or due to GR precession, and ultimately,
by the RR torques themselves. On these long timescales the orbits
evolve in a random walk fashion ($\propto\sqrt{t}$). However, since
$\mathbf{J}$ accumulates a very large {}``mean free path'' over
the coherence time, the resulting random walk in $\mathbf{J}$ proceeds
rapidly on the RR timescale $T_{RR}\!\ll\! T_{R}$. 

\begin{figure}[t]
\noindent \begin{centering}
\includegraphics[width=0.85\textwidth]{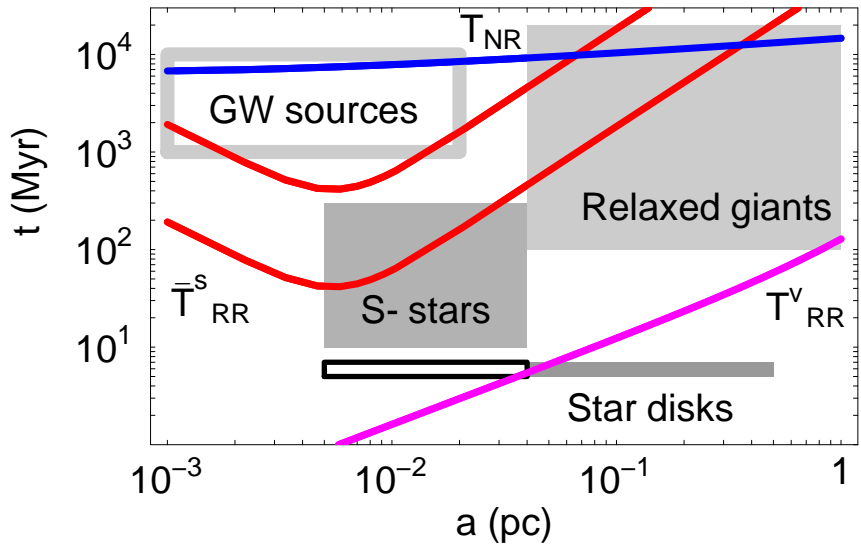}
\par\end{centering}

\caption{\label{f:RR} Possible signature of RR in the GC \citep{hop+06a}.
Significant relaxation is expected where the relaxation times are
shorter than the typical ages of the different dynamical components
(see text). The two lines for scalar RR represent different assumptions
about $M_{\mathrm{eff}}$ (reproduced with permission from the Astrophysical
Journal).}

\end{figure}

There are indications that RR can explain some of the dynamical properties
of the different populations in the GC (Figure \ref{f:RR}). The inner
limit of the stellar disk coincides with the distance where {}``Vector
RR'' (very rapid change in the direction of $\mathbf{J}$ in a spherical
potential) randomizes the inclination of disk orbits on a time-scale
$T_{RR}^{v}\sim QP/N_{\star}^{1/2}$. {}``Scalar RR'' (rapid changes
in the magnitude of $\mathbf{J}$ in a Kepler potential, which falls
with decreasing distance as $T_{RR}^{s}\sim QP$ far from the MBH
due to the enclosed stellar mass, and then rises again as $T_{RR}^{s}\sim6(Q^{2}/\Ns)(r_{g}/a)P$
close to the MBH due to GR precession, where $r_{g}\equiv G\Mbh/c^{2}$
and $a$ is the semi-major axis (sma)), could explain the partially
randomized eccentricities of the S-stars \citep{per+09b}. Vector
RR may also be responsible for randomizing the orbits of the relaxed
giants outside the central $1"$, which are old in term of their main
sequence lifespan, but young compared to the long 2-body relaxation
time.

In a multi-mass population, $T_{RR}\propto M_{\mathrm{eff}}^{-1}$,
where $M_{\mathrm{eff}}\equiv\left.\left\langle \Ms^{2}\right\rangle \right/\left\langle \Ms\right\rangle $,
and so RR is substantially accelerated by mass segregation. It is
noteworthy that the random-walk regime of scalar RR depends only on
$M_{\mathrm{eff}}$, but \emph{not} on $\Ns$. Scalar RR (rapid eccentricity
evolution) can therefore probe mass segregation independently of the
unknown stellar distribution around the MBH. 

\begin{figure}[t]
\noindent \begin{centering}
\begin{tabular}{cc}
\includegraphics[width=0.45\columnwidth]{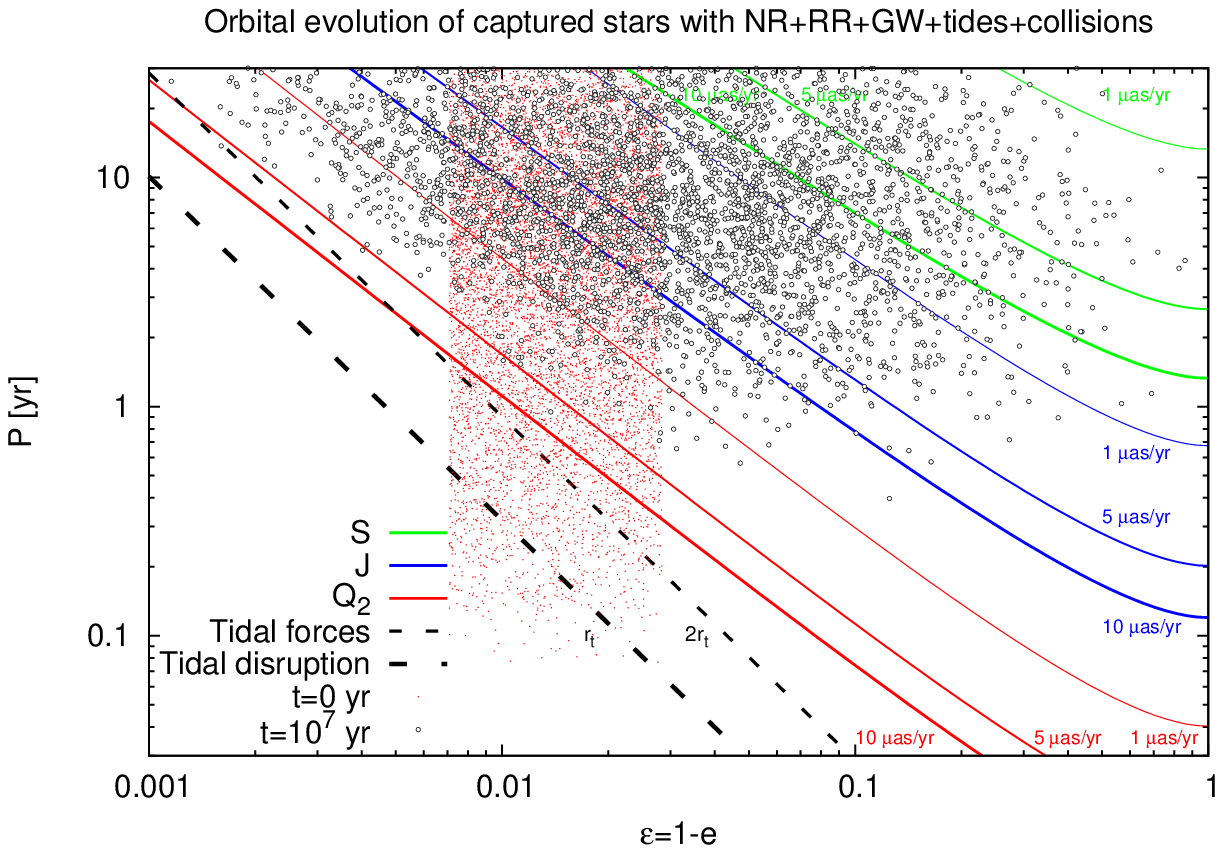} & 
\includegraphics[width=0.45\columnwidth]{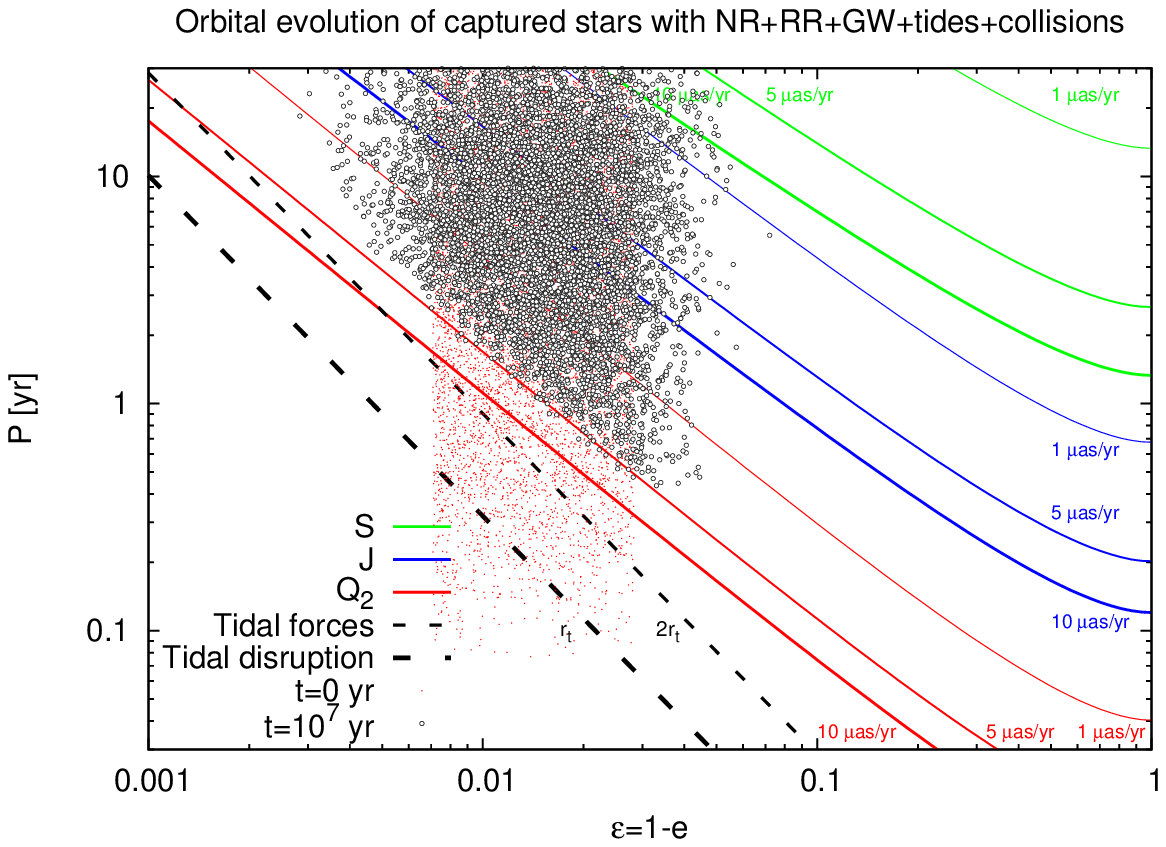}
\tabularnewline
\end{tabular}
\par\end{centering}

\caption{\label{f:GRstars} The post-capture orbital phase-space evolution
of tidally captured stars on relativistic orbits (cf Figure 1 in \citealt{wil09})
(Bar-Or, Alexander \& Perets, in prep.). Left: The initial distribution
(dots) evolves with time by interactions with the dark cusp and the
MBH (points). Right: The initial distribution remains relatively unevolved
in the absence of a cusp. This leads to a higher survival rate, but
also a lower probability of scattering into a short-period relativistic
orbit.}

\end{figure}

A dark cusp will strongly affect the dynamics of stars on relativistic
orbits. These are of particular interest since they can be used to
probe GR in the strong field limit. For example, the precession of
relativistic orbits can be used to test the {}``No Hair Theorem''
\citep{wil09}. RR torques by a dark cusp introduce noise to the GR-driven
orbital evolution, which significantly complicates the detection of
GR effects \citep{mer+10}. Stars on relativistic orbits can not survive
interactions with the MBH and with stellar BHs for more than ${\cal O}(10^{8}\,\mathrm{yr})$,
and therefore they must be continuously replenished, for example by
tidal breakup of incoming binaries (section \ref{s:Sstars}). The
dark cusp will influence the post-capture evolution and survival of
such stars. RR and physical collisions, together with tidal and GW
interactions with the MBH, can populate a fraction of the stars on
very relativistic short-period orbits where GR effects are easier
to detect, at the price of destroying a large fraction of them (Figure
\ref{f:GRstars}).

\section{What is the origin of the stellar disk(s)?}

\label{s:disk}

The disk dynamics of the luminous O-stars in the central $\sim0.5$
pc of the GC \citep{lev+03,pau+06} set strong constraints on possible
formation mechanisms for the young stars. Two leading possibilities
have been considered. The inspiralling cluster scenario \citep{ger01},
and \emph{in situ} SF in a massive fragmenting gas disk (\citealt{pac78};
Levin, Nayakshin, this volume). 

The infalling cluster scenario is disfavored because even a dense
stellar cluster will disintegrate completely by the MBH tidal field
before reaching the central parsec. In order to reach the center,
it must be held together by an IMBH \citep{han+03}, which has to
be an implausibly massive one relative to its cluster mass \citep{gur+05}.
There is to date no compelling evidence for the existence of an IMBH
in the GC (Trippe, this volume). Furthermore, a disintegrating cluster
is expected to leave a tidal tail of stripped stars over a large range
of radii, which are not observed (the distribution of young stars
ends quite sharply at $\sim0.5$ pc, \citealt{pau+06}). 

Observations and modelling currently favor the \emph{in situ} fragmenting
gas disk scenario. The mass function of stars born in a disk is expected
to be top-heavy because the tidal field of the MBH and the disk temperature
imply a higher Jeans mass to begin with, and the massive proto-star
further grows by accretion from the disk \citep[e.g.][]{lev+03}.
There are indeed indications of a top-heavy mass function (\citealt{nay+05a,bar+10};
Bartko, Najarro, this volume). The disk displays marked deviations
from an ideal flat thin disk, as evidenced by the observed non-circular
orbits, warps, wide opening angle, and outlying O-stars \citep{bar+09,Lu+09}.
These deviations are interpreted as post-formation evolution (Bartko,
Madigan, Perets, this volume). Some of the outlying O-stars may be
members of a second, disintegrating counter-rotating disk \citep{pau+06}. 

The opportunity to observe the products of this new channel of SF,
which occurs under conditions that are very different from those of
GMC fragmentation elsewhere in the Galaxy, is of obvious significance
for understanding SF in general.

\section{What is the origin of the S-stars?}

\label{s:Sstars}

The young, seemingly normal main sequence B-stars \citep{ghe+03a,eis+05}
in the inner $\sim1"$ ($\sim0.04$ pc) of the Galactic MBH, the so-called
{}``S-cluster'', pose one of the most intriguing puzzles of the
GC. Simplicity and economy considerations often lead to the natural
assumption that the S-stars are associated with the young stars farther
out. However, there are significant systematic differences between
the S-cluster and the disk stars. Unlike the co-rotating, approximately
circular orbits of the disk stars, the S-stars have random orbital
orientations with even higher orbital eccentricities than expected
in an isotropic distribution \citep{gil+09}. In addition, the brightest
stars in the S-cluster are early B-stars, quite fainter, less massive
and longer lived than the very massive O-stars that define the disk.

Generally, any scenario that postulates a disk origin for the S-stars
(\citealt{mil+04,loe+09,imad+09,gri10}; Madigan, Yelda, this volume),
must also be able to explain this {}``inverse mass segregation''
which concentrates the lower mass stars in the center, while leaving
the more massive stars farther out. No such compelling scenario has
yet been suggested. An alternative explanation that circumvents this
problem is that the B-stars are the most massive survivors of a \emph{previous}
episode of disk fragmentation, which also produced now-dead O-stars
in the S-cluster. However, this then raises the problem why no O-stars
from the \emph{present} disk are seen today in the S-stars cluster. 

An alternative to \emph{in situ }SF scenarios is to assume that the
S-stars migrated to their present position from outside the central
parsec (from the {}``field''), and that they are a distinct population,
unrelated to the disk stars. En-mass migration as part of a stellar
cluster is disfavored by observations, as discussed above (section
\ref{s:disk}). A more promising option is individual capture of stars
by tidal disruption of incoming binaries (3-body exchanges, \citealt{hil88};
Perets, this volume). This process leaves a distinct imprint on the
initial sma and eccentricity distributions of the captured stars.
This can be seen by considering for simplicity equal mass binaries
of mass $2\Ms$ and sma $a_{2}$, which are scattered to the MBH on
a parabolic orbit, and are tidally disrupted at a distance $r_{t}=a_{2}(M_{\bullet}/2\Ms)^{1/3}\,$.
The point of disruption then becomes the periapse of the captured
orbit with sma $a_{1}$ and eccentricity $e_{1}$, $r_{t}=a_{1}(1-e_{1})$.
The orbital energy extracted by the work of the tidal field on the
binary, $\mathrm{d}E\sim[(GM_{\bullet}/r_{t}^{3})a_{2}]r_{t}\sim GM_{\bullet}^{1/3}(2\Ms)^{5/3}/a_{2}$,
is carried by the ejected star, so that the captured orbit has energy
$-\mathrm{d}E$ and $a_{1}=-G\Mbh\Ms/2\mathrm{d}E$. Therefore, the
typical initial capture sma maps the original binary sma, $\left\langle a_{1}\right\rangle \sim(\Mbh/2\Ms)^{2/3}a_{2}$,
and the initial eccentricity is very high and independent of the sma,
$\left\langle e_{1}\right\rangle \sim1-(2\Ms/\Mbh)^{1/3}>0.95$.

\begin{figure}[t]
\noindent \begin{centering}
\includegraphics[width=0.9\columnwidth]{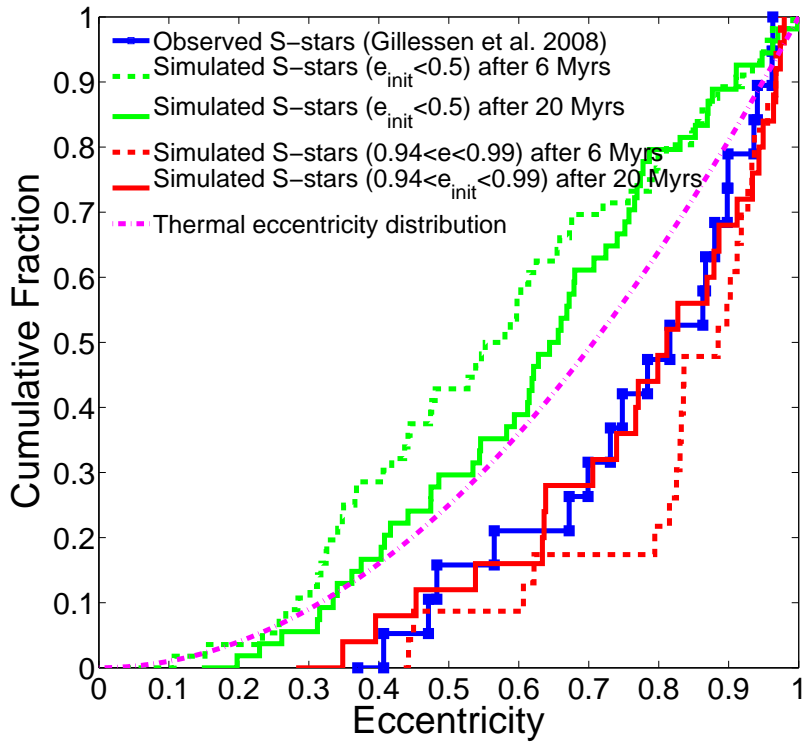}
\par\end{centering}

\caption{\label{f:eSstar} The evolution of S-star eccentricities under different
assumed formation scenarios and timescales, as measured in Newtonian
$N$-body simulations \citep{per+09b}. The best fit is obtained for
a tidal capture scenario with a high initial eccentricity, which evolves
over the typical S-star lifespan of $\sim20$ Myr. Disk origin models
with smaller initial eccentricities do not reach the observed eccentricity
distribution irrespective of assumed age (whether 6 Myr for the disk,
or 20 Myr full lifespan), and instead converge to the isotropic (thermal)
distribution (reproduced with permission from the Astrophysical Journal). }

\end{figure}

Several lines of evidence support the tidal capture scenario. The
luminosity function (LF) of the S-stars is close to the steep (bottom
heavy) universal LF that is observed in the field, and is quite unlike
the flat (top heavy) LF of the disk stars (Bartko, this volume). This
strongly suggests that the S-stars were not formed \emph{in situ},
and are unrelated to the disks. The eccentricity distribution of the
S-stars is more eccentric than in an isotropic distribution \citep{gil+09},
although not as biased to high eccentricities as expected for the
initial post-capture orbits. This is consistent with efficient post-capture
randomization by RR (Figure \ref{f:eSstar}) \citep{per+09b}. The
sma distribution the captured stars is harder to predict, since unlike
the eccentricity distribution, it depends both on the poorly known
sma distribution of the field binaries, and on the details of the
scattering process by the perturbers \citep{per+10}. The tidal capture
mechanism pairs each captured S-star with an ejected hyper velocity
star (HVS). The numbers of observed S-stars and HVSs are consistent.
The tidal capture mechanism also predicts a temporally continuous
distribution of HVSs, which agrees with observations, and a spatially
homogeneous distribution, which may not, albeit with still low statistics
\citep{bro+09} (Brown, Yu, this volume).

Two-body relaxation alone is too slow to deflect massive binaries
from the field at a high enough rate to maintain a steady-state population
of $ $$\sim40$ S-stars. However, such high rates can be driven by
the massive perturbers (GMCs) observed on the $\sim5$--100 pc scale
\citep{per+07}. GMCs are known to play an important role in the dynamics
of the Galactic disk on much larger scales \citep{spi+51}. Their
role in the nuclear dynamics of the Milky Way, suggested by the S-cluster,
implies a significance for the nuclear dynamics of gas rich galaxies
in general. For example, GMCs can drive binary MBHs in post-merger
galaxies to rapid coalescence and the emission of an extremely strong
burst of GW radiation \citep{per+08c}.

\section{Conclusions and summary}

MBHs play many important roles across all fields of astrophysics.
In particular, low-mass MBHs such as the Galactic MBH are the targets
of GW searches by LISA. The dynamics of the GC near the MBH are key
to testing the validity of commonly held assumptions, frequently used
approximations, and theoretical scenarios. Such studies can indicate
whether conclusions that apply to the Galactic MBH can be extrapolated
to other galaxies, and in general, they provide a realistic assessment
of the robustness of dynamical models of galactic nuclei. This short
review focused on four key questions about GC dynamics.
\begin{enumerate}
\item \textbf{Is the stellar system around the MBH relaxed?} The state of
relaxation is directly tied to the shape of the stellar density distribution.
A relaxed old population should exhibit a high density cusp. Conversely,
a flat core or central depletion implies that the system is unrelaxed
and evolving. New observations decisively show that the old red giant
population does \emph{not} have a cusp. If the red giants trace the
entire population, then the GC is unrelaxed, and its dynamical state
reflects some particular initial conditions, which needs not apply
to other galaxies.
\item \textbf{Is there a {}``dark cusp'' around the MBH?} It is possible
that some process selectively destroys red giants, and irrespective
of that, fast relaxation mechanisms could accelerate cusp reformation
even if it was destroyed by a past events. It is therefore of interest
to consider separately the existence of a mass-segregated dark cusp
composed of compact remnants, mostly stellar BHs, and faint low-mass
stars. Stellar black holes are hard to detect directly, but should
have dynamical effects on the orbits of stars. At this time there
is no direct evidence for the existence of a dark cusp in the GC.
\item \textbf{What is the origin of the stellar disk(s)?} Observations favor
\emph{in situ} formation of a top-heavy stellar population in a self-gravitating
fragmenting gas disk. The stellar disk (or disks) show evidence of
substantial post-formation dynamical evolution. 
\item \textbf{What is the origin of the S-stars? }Observations favor an
origin in the field, rather than \emph{in situ} SF. A promising mechanism
is exchange capture of young binaries, which are efficiently deflected
from the field by massive perturbers and then undergo post-capture
orbital evolution.
\end{enumerate}
This work was supported by DIP grant no. 71-0460-0101, ERC Starting
Grant no. 202996 and ISF grant no. 928/06. 

\bibliographystyle{asp2010}

\end{document}